\magnification=\magstephalf
\newbox\SlashedBox 
\def\slashed#1{\setbox\SlashedBox=\hbox{#1}
\hbox to 0pt{\hbox to 1\wd\SlashedBox{\hfil/\hfil}\hss}{#1}}
\def\hboxtosizeof#1#2{\setbox\SlashedBox=\hbox{#1}
\hbox to 1\wd\SlashedBox{#2}}

\def\mathslashed#1{\setbox\SlashedBox=\hbox{$#1$}
\hbox to 0pt{\hbox to 1\wd\SlashedBox{\hfil/\hfil}\hss}#1}

\def\ifsmall{\iffalse}  
\def\titlepagefont{}  

\def\DefineTeXgraphics{%
\special{ps::[global] /TeXgraphics { } def}}  

\def\today{\ifcase\month\or January\or February\or March\or April\or May
\or June\or July\or August\or September\or October\or November\or
December\fi\space\number\day, \number\year}
\def\eatPrefix19{}
\def\Year{\expandafter\eatPrefix\the\year}
\newcount\hours \newcount\minutes
\def\monthname{\ifcase\month\or
January\or February\or March\or April\or May\or June\or July\or
August\or September\or October\or November\or December\fi}
\def\shortmonthname{\ifcase\month\or
Jan\or Feb\or Mar\or Apr\or May\or Jun\or Jul\or
Aug\or Sep\or Oct\or Nov\or Dec\fi}

\def\TimeStamp{\hours\the\time\divide\hours by60%
\minutes -\the\time\divide\minutes by60\multiply\minutes by60%
\advance\minutes by\the\time%
${\rm \shortmonthname}\cdot\if\day<10{}0\fi\the\day\cdot\the\year%
\qquad\the\hours:\if\minutes<10{}0\fi\the\minutes$}




\def\Title#1{%
\vskip 1in{\titlefont\centerline{#1}}\vskip .5in}
 
\def\Date#1{\leftline{#1}\tenrm\supereject%
\global\hsize=\hsbody\global\hoffset=\hbodyoffset%
\footline={\hss\tenrm\folio\hss}}

\newif\ifdraftmode
\newif\ifleftlabels  

\def\nolabels{\def\wrlabeL##1{}\def\eqlabeL##1{}\def\reflabeL##1{}}
\def\writelabels{\def\wrlabeL##1{\leavevmode\vadjust{\rlap{\smash%
{\line{{\escapechar=` \hfill\rlap{\sevenrm\hskip.03in\string##1}}}}}}}%
\def\eqlabeL##1{{\escapechar-1\rlap{\sevenrm\hskip.05in\string##1}}}%
\def\reflabeL##1{\noexpand\rlap{\noexpand\sevenrm[\string##1]}}}
\def\writeleftlabels{\def\wrlabeL##1{\leavevmode\vadjust{\rlap{\smash%
{\line{{\escapechar=` \hfill\rlap{\sevenrm\hskip.03in\string##1}}}}}}}%
\def\eqlabeL##1{{\escapechar-1%
\rlap{\sixrm\hskip.05in\string##1}%
\llap{\sevenrm\string##1\hskip.03in\hbox to \hsize{}}}}%
\def\reflabeL##1{\noexpand\rlap{\noexpand\sevenrm[\string##1]}}}
\nolabels

\input hyperbasics.tex

\newdimen\fullhsize
\newdimen\hstitle
\hstitle=\hsize 
\newdimen\hsbody
\hsbody=\hsize 
\newdimen\hbodyoffset
\hbodyoffset=\hoffset 
\newbox\leftpage
\def\abstract#1{#1}
\def\rotated{\special{ps: landscape}
\magnification=1000  
\baselineskip=14pt
\global\hstitle=9truein\global\hsbody=4.75truein
\global\vsize=7truein\global\voffset=-.31truein
\global\hoffset=-0.54in\global\hbodyoffset=-.54truein
\global\fullhsize=10truein
\def\DefineTeXgraphics{%
\special{ps::[global] 
/TeXgraphics {currentpoint translate 0.7 0.7 scale
              -80 0.72 mul -1000 0.72 mul translate} def}}
\let\lr=L
\def\ifsmall{\iftrue}
\def\titlepagefont{\twelvepoint}
\trueseventeenpoint
\def\almostshipout##1{\if L\lr \count1=1
      \global\setbox\leftpage=##1 \global\let\lr=R
   \else \count1=2
      \shipout\vbox{\hbox to\fullhsize{\box\leftpage\hfil##1}}
      \global\let\lr=L\fi}

\output={\ifnum\count0=1 
 \shipout\vbox{\hbox to \fullhsize{\hfill\pagebody\hfill}}\advancepageno
 \else
 \almostshipout{\leftline{\vbox{\pagebody\makefootline}}}\advancepageno 
 \fi}

\def\abstract##1{{\leftskip=1.5in\rightskip=1.5in ##1\par}} }

\def\linemessage#1{\immediate\write16{#1}}

\global\newcount\secno \global\secno=0
\global\newcount\appno \global\appno=0
\global\newcount\meqno \global\meqno=1
\global\newcount\subsecno \global\subsecno=0
\global\newcount\figno \global\figno=0

\newif\ifAnyCounterChanged
\let\terminator=\relax
\def\normalize#1{\ifx#1\terminator\let\next=\relax\else%
\if#1i\aftergroup i\else\if#1v\aftergroup v\else\if#1x\aftergroup x%
\else\if#1l\aftergroup l\else\if#1c\aftergroup c\else%
\if#1m\aftergroup m\else%
\if#1I\aftergroup I\else\if#1V\aftergroup V\else\if#1X\aftergroup X%
\else\if#1L\aftergroup L\else\if#1C\aftergroup C\else%
\if#1M\aftergroup M\else\aftergroup#1\fi\fi\fi\fi\fi\fi\fi\fi\fi\fi\fi\fi%
\let\next=\normalize\fi%
\next}
\def\makeNormal#1#2{\def\doNormalDef{\edef#1}\begingroup%
\aftergroup\doNormalDef\aftergroup{\normalize#2\terminator\aftergroup}%
\endgroup}

\def\warnIfChanged#1#2{%
\ifundef#1
\else\begingroup%
\edef\oldDefinitionOfCounter{#1}\edef\newDefinitionOfCounter{#2}%
\ifx\oldDefinitionOfCounter\newDefinitionOfCounter%
\else%
\linemessage{Warning: definition of \noexpand#1 has changed.}%
\global\AnyCounterChangedtrue\fi\endgroup\fi}

\def\Section#1{\global\advance\secno by1\relax\global\meqno=1%
\global\subsecno=0%
\bigbreak\bigskip
\centerline{\twelvepoint \bf %
\the\secno. #1}%
\par\nobreak\medskip\nobreak}
\def\tagsection#1{%
\warnIfChanged#1{\the\secno}%
\xdef#1{\the\secno}%
\ifWritingAuxFile\immediate\write\auxfile{\noexpand\xdef\noexpand#1{#1}}\fi%
}
\def\section{\Section}
\def\Subsection#1{\global\advance\subsecno by1\relax\medskip %
\leftline{\bf\the\secno.\the\subsecno\ #1}%
\par\nobreak\smallskip\nobreak}
\def\tagsubsection#1{%
\warnIfChanged#1{\the\secno.\the\subsecno}%
\xdef#1{\the\secno.\the\subsecno}%
\ifWritingAuxFile\immediate\write\auxfile{\noexpand\xdef\noexpand#1{#1}}\fi%
}

\def\subsection{\Subsection}

\def\romappno{\uppercase\expandafter{\romannumeral\appno}}
\def\makeNormalizedRomappno{%
\expandafter\makeNormal\expandafter\normalizedromappno%
\expandafter{\romannumeral\appno}%
\edef\normalizedromappno{\uppercase{\normalizedromappno}}}
\def\Appendix#1{\global\advance\appno by1\relax\global\meqno=1\global\secno=0%
\global\subsecno=0%
\bigbreak\bigskip
\centerline{\twelvepoint \bf Appendix %
\romappno. #1}%
\par\nobreak\medskip\nobreak}
\def\tagappendix#1{\makeNormalizedRomappno%
\warnIfChanged#1{\normalizedromappno}%
\xdef#1{\normalizedromappno}%
\ifWritingAuxFile\immediate\write\auxfile{\noexpand\xdef\noexpand#1{#1}}\fi%
}
\def\appendix{\Appendix}
\def\Subappendix#1{\global\advance\subsecno by1\relax\medskip %
\leftline{\bf\romappno.\the\subsecno\ #1}%
\par\nobreak\smallskip\nobreak}
\def\tagsubappendix#1{\makeNormalizedRomappno%
\warnIfChanged#1{\normalizedromappno.\the\subsecno}%
\xdef#1{\normalizedromappno.\the\subsecno}%
\ifWritingAuxFile\immediate\write\auxfile{\noexpand\xdef\noexpand#1{#1}}\fi%
}

\def\eqn#1{\makeNormalizedRomappno%
\ifnum\secno>0%
  \warnIfChanged#1{\the\secno.\the\meqno}%
  \eqno(\the\secno.\the\meqno)\xdef#1{\the\secno.\the\meqno}%
     \global\advance\meqno by1
\else\ifnum\appno>0%
  \warnIfChanged#1{\normalizedromappno.\the\meqno}%
  \eqno({\rm\romappno}.\the\meqno)%
      \xdef#1{\normalizedromappno.\the\meqno}%
     \global\advance\meqno by1
\else%
  \warnIfChanged#1{\the\meqno}%
  \eqno(\the\meqno)\xdef#1{\the\meqno}%
     \global\advance\meqno by1
\fi\fi%
\eqlabeL#1%
\ifWritingAuxFile\immediate\write\auxfile{\noexpand\xdef\noexpand#1{#1}}\fi%
}
\def\defeqn#1{\makeNormalizedRomappno%
\ifnum\secno>0%
  \warnIfChanged#1{\the\secno.\the\meqno}%
  \xdef#1{\the\secno.\the\meqno}%
     \global\advance\meqno by1
\else\ifnum\appno>0%
  \warnIfChanged#1{\normalizedromappno.\the\meqno}%
  \xdef#1{\normalizedromappno.\the\meqno}%
     \global\advance\meqno by1
\else%
  \warnIfChanged#1{\the\meqno}%
  \xdef#1{\the\meqno}%
     \global\advance\meqno by1
\fi\fi%
\eqlabeL#1%
\ifWritingAuxFile\immediate\write\auxfile{\noexpand\xdef\noexpand#1{#1}}\fi%
}
\def\anoneqn{\makeNormalizedRomappno%
\ifnum\secno>0
  \eqno(\the\secno.\the\meqno)%
     \global\advance\meqno by1
\else\ifnum\appno>0
  \eqno({\rm\normalizedromappno}.\the\meqno)%
     \global\advance\meqno by1
\else
  \eqno(\the\meqno)%
     \global\advance\meqno by1
\fi\fi%
}
\def\mfig#1#2{\ifx#20
\else\global\advance\figno by1%
\relax#1\the\figno%
\warnIfChanged#2{\the\figno}%
\xdef#2{\the\figno}%
\reflabeL#2%
\ifWritingAuxFile\immediate\write\auxfile{\noexpand\xdef\noexpand#2{#2}}\fi\fi%
}

\catcode`@=11 

\newif\ifFiguresInText\FiguresInTexttrue
\newif\if@FigureFileCreated
\newwrite\capfile
\newwrite\figfile

\newif\ifcaption
\captiontrue
\def\captionsize{\tenrm}
\def\PlaceTextFigure#1#2#3#4{%
\vskip 0.5truein%
#3\hfil\epsfbox{#4}\hfil\break%
\ifcaption\hfil\vbox{\captionsize Figure #1. #2}\hfil\fi%
\vskip10pt}
\def\PlaceEndFigure#1#2{%
\epsfxsize=\hsize\epsfbox{#2}\vfill\centerline{Figure #1.}\eject}

\def\LoadFigure#1#2#3#4{%
\ifundef#1{\phantom{\mfig{}#1}}\else
\fi%
\ifFiguresInText
\PlaceTextFigure{#1}{#2}{#3}{#4}%
\else
\if@FigureFileCreated\else%
\immediate\openout\capfile=\jobname.caps%
\immediate\openout\figfile=\jobname.figs%
@FigureFileCreatedtrue\fi%
\immediate\write\capfile{\noexpand\item{Figure \noexpand#1.\ }{#2}\vskip10pt}%
\immediate\write\figfile{\noexpand\PlaceEndFigure\noexpand#1{\noexpand#4}}%
\fi}

\def\listfigs{\ifFiguresInText\else%
\vfill\eject\immediate\closeout\capfile
\immediate\closeout\figfile%
\centerline{{\bf Figures}}\bigskip\frenchspacing%
\catcode`@=11 
\def\captionsize{\tenrm}
\input \jobname.caps\vfill\eject\nonfrenchspacing%
\catcode`\@=\active
\catcode`@=12  
\input\jobname.figs\fi}

\font\ninerm=cmr9
\font\eightrm=cmr8
\font\sixrm=cmr6

\def\loadtrueseventeenpoint{
 \font\seventeenrm=cmr10 at 17.28truept
 \font\seventeeni=cmmi10 at 17.28truept
 \font\seventeenbf=cmbx10 at 17.28truept
 \font\seventeenit=cmti10 at 17.28truept
 \font\seventeensl=cmsl10 at 17.28truept
 \font\seventeensy=cmsy10 at 17.28truept
}
\def\loadfourteenpoint{
\font\fourteenrm=cmr10 at 14.4pt
\font\fourteeni=cmmi10 at 14.4pt
\font\fourteenit=cmti10 at 14.4pt
\font\fourteensl=cmsl10 at 14.4pt
\font\fourteensy=cmsy10 at 14.4pt
\font\fourteenbf=cmbx10 at 14.4pt
}
\def\loadtruetwelvepoint{
\font\twelverm=cmr10 at 12truept
\font\twelvei=cmmi10 at 12truept
\font\twelveit=cmti10 at 12truept
\font\twelvesl=cmsl10 at 12truept
\font\twelvesy=cmsy10 at 12truept
\font\twelvebf=cmbx10 at 12truept
}

\font\ninei=cmmi9
\font\eighti=cmmi8
\font\sixi=cmmi6
\skewchar\ninei='177 \skewchar\eighti='177 \skewchar\sixi='177

\font\ninesy=cmsy9
\font\eightsy=cmsy8
\font\sixsy=cmsy6
\skewchar\ninesy='60 \skewchar\eightsy='60 \skewchar\sixsy='60

\font\ninebf=cmbx9
\font\eightbf=cmbx8
\font\sixbf=cmbx6

\font\ninett=cmtt9
\font\eighttt=cmtt8

\hyphenchar\tentt=-1 
\hyphenchar\ninett=-1
\hyphenchar\eighttt=-1         

\font\ninesl=cmsl9
\font\eightsl=cmsl8

\font\nineit=cmti9
\font\eightit=cmti8
\font\sevenit=cmti7

\scriptfont\itfam=\sevenit

                      
\newskip\ttglue
\def\tenpoint{\def\rm{\fam0\tenrm}%
  \textfont0=\tenrm \scriptfont0=\sevenrm \scriptscriptfont0=\fiverm
  \textfont1=\teni \scriptfont1=\seveni \scriptscriptfont1=\fivei
  \textfont2=\tensy \scriptfont2=\sevensy \scriptscriptfont2=\fivesy
  \textfont3=\tenex \scriptfont3=\tenex \scriptscriptfont3=\tenex
  \def\it{\fam\itfam\tenit}%
      \textfont\itfam=\tenit\scriptfont\itfam=\sevenit
  \def\sl{\fam\slfam\tensl}\textfont\slfam=\tensl
  \def\bf{\fam\bffam\tenbf}\textfont\bffam=\tenbf \scriptfont\bffam=\sevenbf
  \scriptscriptfont\bffam=\fivebf
  \normalbaselineskip=12pt
  \let\sc=\eightrm
  \let\big=\tenbig
  \setbox\strutbox=\hbox{\vrule height8.5pt depth3.5pt width\z@}%
  \normalbaselines\rm}

\def\twelvepoint{\def\rm{\fam0\twelverm}%
  \textfont0=\twelverm \scriptfont0=\ninerm \scriptscriptfont0=\sevenrm
  \textfont1=\twelvei \scriptfont1=\ninei \scriptscriptfont1=\seveni
  \textfont2=\twelvesy \scriptfont2=\ninesy \scriptscriptfont2=\sevensy
  \textfont3=\tenex \scriptfont3=\tenex \scriptscriptfont3=\tenex
  \def\it{\fam\itfam\twelveit}\textfont\itfam=\twelveit
  \def\sl{\fam\slfam\twelvesl}\textfont\slfam=\twelvesl
  \def\bf{\fam\bffam\twelvebf}\textfont\bffam=\twelvebf%
  \scriptfont\bffam=\ninebf
  \scriptscriptfont\bffam=\sevenbf
  \normalbaselineskip=12pt
  \let\sc=\eightrm
  \let\big=\tenbig
  \setbox\strutbox=\hbox{\vrule height8.5pt depth3.5pt width\z@}%
  \normalbaselines\rm}

\def\fourteenpoint{\def\rm{\fam0\fourteenrm}%
  \textfont0=\fourteenrm \scriptfont0=\tenrm \scriptscriptfont0=\sevenrm
  \textfont1=\fourteeni \scriptfont1=\teni \scriptscriptfont1=\seveni
  \textfont2=\fourteensy \scriptfont2=\tensy \scriptscriptfont2=\sevensy
  \textfont3=\tenex \scriptfont3=\tenex \scriptscriptfont3=\tenex
  \def\it{\fam\itfam\fourteenit}\textfont\itfam=\fourteenit
  \def\sl{\fam\slfam\fourteensl}\textfont\slfam=\fourteensl
  \def\bf{\fam\bffam\fourteenbf}\textfont\bffam=\fourteenbf%
  \scriptfont\bffam=\tenbf
  \scriptscriptfont\bffam=\sevenbf
  \normalbaselineskip=17pt
  \let\sc=\elevenrm
  \let\big=\tenbig                                          
  \setbox\strutbox=\hbox{\vrule height8.5pt depth3.5pt width\z@}%
  \normalbaselines\rm}

\def\seventeenpoint{\def\rm{\fam0\seventeenrm}%
  \textfont0=\seventeenrm \scriptfont0=\fourteenrm \scriptscriptfont0=\tenrm
  \textfont1=\seventeeni \scriptfont1=\fourteeni \scriptscriptfont1=\teni
  \textfont2=\seventeensy \scriptfont2=\fourteensy \scriptscriptfont2=\tensy
  \textfont3=\tenex \scriptfont3=\tenex \scriptscriptfont3=\tenex
  \def\it{\fam\itfam\seventeenit}\textfont\itfam=\seventeenit
  \def\sl{\fam\slfam\seventeensl}\textfont\slfam=\seventeensl
  \def\bf{\fam\bffam\seventeenbf}\textfont\bffam=\seventeenbf%
  \scriptfont\bffam=\fourteenbf
  \scriptscriptfont\bffam=\twelvebf
  \normalbaselineskip=21pt
  \let\sc=\fourteenrm
  \let\big=\tenbig                                          
  \setbox\strutbox=\hbox{\vrule height 12pt depth 6pt width\z@}%
  \normalbaselines\rm}

\def\ninepoint{\def\rm{\fam0\ninerm}%
  \textfont0=\ninerm \scriptfont0=\sixrm \scriptscriptfont0=\fiverm
  \textfont1=\ninei \scriptfont1=\sixi \scriptscriptfont1=\fivei
  \textfont2=\ninesy \scriptfont2=\sixsy \scriptscriptfont2=\fivesy
  \textfont3=\tenex \scriptfont3=\tenex \scriptscriptfont3=\tenex
  \def\it{\fam\itfam\nineit}\textfont\itfam=\nineit
  \def\sl{\fam\slfam\ninesl}\textfont\slfam=\ninesl
  \def\bf{\fam\bffam\ninebf}\textfont\bffam=\ninebf \scriptfont\bffam=\sixbf
  \scriptscriptfont\bffam=\fivebf
  \normalbaselineskip=11pt
  \let\sc=\sevenrm
  \let\big=\ninebig
  \setbox\strutbox=\hbox{\vrule height8pt depth3pt width\z@}%
  \normalbaselines\rm}

\def\eightpoint{\def\rm{\fam0\eightrm}%
  \textfont0=\eightrm \scriptfont0=\sixrm \scriptscriptfont0=\fiverm%
  \textfont1=\eighti \scriptfont1=\sixi \scriptscriptfont1=\fivei%
  \textfont2=\eightsy \scriptfont2=\sixsy \scriptscriptfont2=\fivesy%
  \textfont3=\tenex \scriptfont3=\tenex \scriptscriptfont3=\tenex%
  \def\it{\fam\itfam\eightit}\textfont\itfam=\eightit%
  \def\sl{\fam\slfam\eightsl}\textfont\slfam=\eightsl%
  \def\bf{\fam\bffam\eightbf}\textfont\bffam=\eightbf \scriptfont\bffam=\sixbf%
  \scriptscriptfont\bffam=\fivebf%
  \normalbaselineskip=9pt%
  \let\sc=\sixrm%
  \let\big=\eightbig%
  \setbox\strutbox=\hbox{\vrule height7pt depth2pt width\z@}%
  \normalbaselines\rm}

\def\tenbig#1{{\hbox{$\left#1\vbox to8.5pt{}\right.\n@space$}}}
\def\ninebig#1{{\hbox{$\textfont0=\tenrm\textfont2=\tensy
  \left#1\vbox to7.25pt{}\right.\n@space$}}}
\def\eightbig#1{{\hbox{$\textfont0=\ninerm\textfont2=\ninesy
  \left#1\vbox to6.5pt{}\right.\n@space$}}}

\def\footnote#1{\edef\@sf{\spacefactor\the\spacefactor}#1\@sf
      \insert\footins\bgroup\eightpoint
      \interlinepenalty100 \let\par=\endgraf
        \leftskip=\z@skip \rightskip=\z@skip
        \splittopskip=10pt plus 1pt minus 1pt \floatingpenalty=20000
        \smallskip\item{#1}\bgroup\strut\aftergroup\@foot\let\next}
\skip\footins=12pt plus 2pt minus 4pt 
\dimen\footins=30pc 

\newinsert\margin
\dimen\margin=\maxdimen
\def\titlefont{\seventeenpoint}
\loadtruetwelvepoint 
\loadtrueseventeenpoint

\def\eatOne#1{}
\def\ifundef#1{\expandafter\ifx%
\csname\expandafter\eatOne\string#1\endcsname\relax}
\def\notTrue{\iffalse}\def\isTrue{\iftrue}
\def\ifdef#1{{\ifundef#1%
\aftergroup\notTrue\else\aftergroup\isTrue\fi}}
\def\use#1{\ifundef#1\linemessage{Warning: \string#1 is undefined.}%
{\tt \string#1}\else#1\fi}



%
\catcode`"=11
\let\quote="
\catcode`"=12
\chardef\foo="22
\global\newcount\refno \global\refno=1
\newwrite\rfile
\newlinechar=`\^^J
\def\@ref#1#2{\the\refno\n@ref#1{#2}}
\def\h@ref#1#2#3{\href{#3}{\the\refno}\n@ref#1{#2}}
\def\n@ref#1#2{\xdef#1{\the\refno}%
\ifnum\refno=1\immediate\openout\rfile=\jobname.refs\fi%
\immediate\write\rfile{\noexpand\item{[\noexpand#1]\ }#2.}%
\global\advance\refno by1}
\def\nref{\n@ref} 
\def\ref{\@ref}   
\def\hrref{\h@ref}
\def\lref#1#2{\the\refno\xdef#1{\the\refno}%
\ifnum\refno=1\immediate\openout\rfile=\jobname.refs\fi%
\immediate\write\rfile{\noexpand\item{[\noexpand#1]\ }#2\semi}%
\global\advance\refno by1}
\def\cref#1{\immediate\write\rfile{#1\semi}}

\def\preref#1#2{\gdef#1{\@ref#1{#2}}}

\def\semi{;\hfil\noexpand\break}

\def\listrefs{\vfill\eject\immediate\closeout\rfile
\centerline{{\bf References}}\bigskip\frenchspacing%
\input \jobname.refs\vfill\eject\nonfrenchspacing}

\def\inputAuxIfPresent#1{\immediate\openin1=#1
\ifeof1\message{No file \auxfileName; I'll create one.
}\else\closein1\relax\input\auxfileName\fi%
}
\def\NPB{Nucl.\ Phys.\ B}




\newif\ifWritingAuxFile
\newwrite\auxfile
\def\SetUpAuxFile{%
\xdef\auxfileName{\jobname.aux}%
\inputAuxIfPresent{\auxfileName}%
\WritingAuxFiletrue%
\immediate\openout\auxfile=\auxfileName}

\def\L{\left(}\def\R{\right)}
\def\LP{\left.}\def\RP{\right.}
\def\LB{\left[}\def\RB{\right]}

\def\RV{\right|}

\def\bye{\par\vfill\supereject%
\ifAnyCounterChanged\linemessage{
Some counters have changed.  Re-run tex to fix them up.}\fi%
\end}

\catcode`\@=\active
\catcode`@=12  
\catcode`\"=\active

\def\Tr{\mathop{\rm Tr}\nolimits}
\def\pol{\varepsilon}

\def\spa#1.#2{\left\langle#1\,#2\right\rangle}
\def\spb#1.#2{\left[#1\,#2\right]}
\SetUpAuxFile
\hfuzz 20pt
\overfullrule 0pt

\def\e{\epsilon}
\def\tree{{\rm tree}}
\def\Split{\mathop{\rm Split}\nolimits}
\def\Soft{\mathop{\rm Soft}\nolimits}
\def\Sing{\mathop{\rm Ant}\nolimits}
\def\tcdot{\mskip -1mu\cdot\mskip-1mu}

\def\llongrightarrow{%
\relbar\mskip-0.5mu\joinrel\mskip-0.5mu\relbar\mskip-0.5mu\joinrel\longrightarrow}
\def\inlimit^#1{\buildrel#1\over\llongrightarrow}

\loadfourteenpoint


\noindent\nopagenumbers
[hep-ph/9710213] \hfill{Saclay/SPhT--T97/109v2}

\leftlabelstrue
\vskip -1.0 in
\Title{Antenna Factorization of Gauge-Theory Amplitudes}

\baselineskip17truept
\centerline{David A. Kosower}
\baselineskip12truept
\centerline{\it Service de Physique Th\'eorique${}^{\natural}$}
\centerline{\it Centre d'Etudes de Saclay}
\centerline{\it F-91191 Gif-sur-Yvette cedex, France}
\centerline{\tt kosower@spht.saclay.cea.fr}

\vskip 0.2in\baselineskip13truept

\vskip 0.5truein
\centerline{\bf Abstract}

{\narrower 
I derive a single factorization formula which summarizes all soft and
collinear divergences of a tree-level gauge theory amplitude.
The singular factor squared is in a certain sense the 
generalization of the Catani--Seymour dipole factorization formula.

}
\vskip 0.3truein

\centerline{\it Submitted to Physical Review D}

\vfill
\vskip 0.1in
\noindent\hrule width 3.6in\hfil\break
\noindent
${}^{\natural}$Laboratory of the
{\it Direction des Sciences de la Mati\`ere\/}
of the {\it Commissariat \`a l'Energie Atomique\/} of France.\hfil\break

\Date{}

\line{}

\baselineskip17pt
%

\preref\Color{%
F. A. Berends and W. T. Giele,
Nucl.\ Phys.\ B294:700 (1987)\semi
D. A.\ Kosower, B.-H.\ Lee and V. P. Nair, Phys.\ Lett.\ 201B:85 (1988)\semi
M.\ Mangano, S. Parke and Z.\ Xu, Nucl.\ Phys.\ B298:653 (1988)\semi
Z. Bern and D. A.\ Kosower, Nucl.\ Phys.\ B362:389 (1991)}

\preref\GG{W.T.\ Giele and E.W.N.\ Glover,
Phys.\ Rev.\ D46:1980 (1992)}

\preref\Recurrence{F. A. Berends and W. T. Giele, Nucl.\ Phys.\ B306:759 (1988)\semi
D. A. Kosower, Nucl.\ Phys.\ B335:23 (1990)}

\preref\ManganoReview{%
M. Mangano and S.J. Parke, Phys.\ Rep.\ 200:301 (1991)}

\preref\CS{%
S. Catani and M. Seymour, Phys.\ Lett.\ B378:287 (1996) [hep-ph/9602277];
Nucl.\ Phys.\ B485:291 (1997) [hep-ph/9605323]}

\preref\SoftGluonReview{A. Bassetto, M. Ciafaloni, and G. Marchesini, 
 Phys.\ Rep.\ 100:201 (1983)}
\preref\AP{G. Altarelli and G. Parisi, Nucl.\ Phys.\ B126:298 (1977)}
\preref\FKS{S. Frixione, Z. Kunszt, and A. Signer, 
   Nucl.\ Phys.\ B467:399 (1996) [hep-ph/9512328]}

\preref\Byckling{E. Byckling and K. Kajantie, {\it Particle Kinematics\/}
(Wiley, 1973)}

\preref\KobaNielsenR{Z. Koba and H. B. Nielsen, Nucl.\ Phys.\ B10:633 (1969)\semi
J. H. Schwarz, Phys.\ Rep.\ 89:223 (1982)}

\preref\CDR{J. C.\ Collins, {\it Renormalization} (Cambridge University Press, 1984)}
\preref\HV{G. 't Hooft and M. Veltman, Nucl.\ Phys.\ B44:189 (1972)}
\preref\DimRed{W. Siegel, Phys.\ Lett.\ 84B:193 (1979)\semi
D.M.\ Capper, D.R.T.\ Jones and P. van Nieuwenhuizen, Nucl.\ Phys.\
B167:479 (1980)\semi
L.V.\ Avdeev and A.A.\ Vladimirov, Nucl.\ Phys.\ B219:262 (1983)}
\preref\FDHS{Z. Bern and D. A.\ Kosower, \NPB 379:451 (1992)}

\section{Introduction}
\vskip 5pt

The factorization properties of gauge-theory amplitudes in infrared-singular
limits have played an important role not only in our conceptual
understanding of infrared safety, but also in explicit calculations at next-to-leading
order in perturbative QCD.  
These infrared singularities may be classified into two types:
soft and collinear.  The former arises when a gluon four-momentum
vanishes, $k_s\rightarrow 0$; the latter when the momenta of
two massless particles become proportional, $k_a\rightarrow z (k_a+k_b)$,
$k_b\rightarrow (1-z) (k_a+k_b)$.  The properties of non-Abelian 
gauge-theory amplitudes in these limits are easiest to understand
in the context of a color decomposition~[\use\Color].  For 
tree-level all-gluon amplitudes in an $SU(N)$ gauge theory 
the color decomposition has the form,
$$
{\cal A}_n^\tree(\{k_i,\lambda_i,a_i\}) = 
\sum_{\sigma \in S_n/Z_n} \Tr(T^{a_{\sigma(1)}}\cdots T^{a_{\sigma(n)}})\,
A_n^\tree(\sigma(1^{\lambda_1},\ldots,n^{\lambda_n}))\,,
\eqn\ColorDecomposition$$
where $S_n/Z_n$ is the group of non-cyclic permutations
on $n$ symbols, and $j^{\lambda_j}$ denotes the $j$-th momentum
and helicity.  As is by now standard,
I use the normalization $\Tr(T^a T^b) = \delta^{ab}$.
One can write analogous formul\ae\ for amplitudes
with quark-antiquark pairs or uncolored external lines.
The color-ordered or partial amplitude $A_n$ is gauge invariant, and has
simple factorization properties in both the soft and collinear limits,
$$
\eqalign{
A_{n}^\tree(\ldots,a^{\lambda_a},b^{\lambda_b},\ldots)
 &\inlimit^{a \parallel b}
\sum_{\lambda=\pm}  
  \Split^{\rm tree}_{-\lambda}(a^{\lambda_a},b^{\lambda_b})\,
      A_{n-1}^\tree(\ldots,(a+b)^\lambda,\ldots)\,,\cr
A_{n}^\tree(\ldots,a,s^{\lambda_s},b,\ldots)
 &\inlimit^{k_s\rightarrow 0}
  \Soft^\tree(a,s^{\lambda_s},b)\,
      A_{n-1}^\tree(\ldots,a,b,\ldots)\,.\cr
}\eqn\Factorization
$$

The collinear splitting amplitude $\Split^{\rm tree}$, 
squared and summed over helicities,
gives the usual unpolarized Altarelli--Parisi splitting function~[\use\AP].
While the complete amplitude also factorizes in the collinear limit,
the same is not true of the soft limit; the eikonal factors $\Soft^\tree$
get tangled up with the color structure.  It is for this reason
that the color decomposition is useful.  Its use, and the
simple factorization properties of the partial amplitudes,
allowed Giele and Glover, in their pioneering paper~[\use\GG], to derive 
simple and universal functions summarizing the integration over
soft and collinear radiation.  Of course, it is possible, as
was done by later authors~[\use\FKS,\use\CS], to square the
amplitude first, and tangle up the color factors with the eikonal
factors, thereby recovering earlier forms of these two
factorizations~[\use\AP,\use\SoftGluonReview],
but this obscures the
clean structure in eqn.~(\use\Factorization).

The color decomposition (\use\ColorDecomposition) is also the form
in which one writes tree-level amplitudes in an open string theory, and the
gauge-theory partial amplitudes may be computed as the infinite-tension
limit of the corresponding string amplitudes.  These are given
by the Koba--Nielsen formula~[\use\KobaNielsenR],
$$\eqalign{
A_n^{\rm KN}&=
\tau^{n-4} \,2^{-n/2} \mskip -50mu
\mathop{\int}_{x_a<x_1<x_b\cdots<x_{n-2}} \prod_{i=1}^{n-3} dx_i
   \,(x_b-x_a)(x_{n-2}-x_a)(x_{n-2}-x_b)\;\cr
&\hskip 20mm\times
\prod_{i\neq j \in E}
       |x_j-x_i|^{\tau^2 k_i\cdot k_j/2}\;
\LP\exp\LB \sum_{i\neq j} {1\over2} {\pol_i\tcdot\pol_j\over (x_j-x_i)^2}
                         + {\tau k_i\tcdot \pol_j\over (x_i-x_j)}\RB
  \RV_{\rm multilinear}\,,\cr
}\eqn\KobaNielsen$$
which I have written in a form that will be useful for the
derivation to be presented in this paper.
In this equation, $\{k_i,\pol_i\}$ are the 
momenta and polarization vectors of the gluons; $E$ denotes the 
set of all external legs, $\{a,1,b,2,\ldots,n-2\}$; 
$\tau$ is the square
root of the inverse string tension; and the subscript `multilinear' tells us
to extract the terms in the exponential linear in each of the polarization
vectors.  Three of the $x_i$ may be fixed at will; it will be convenient
to choose $x_a=0$ and $x_b=1$.
The gauge theory limit is given by $\tau\rightarrow 0$.
This formula gives an explicit form for the $n$-gluon amplitude, 
and examining its limits is thus a convenient method of 
deriving~[\use\ManganoReview]
factorization formul\ae\ such as those in eqn.~(\use\Factorization).

More recently, Catani and Seymour~[\use\CS] wrote down a {\it dipole\/} factorization
formula which gives a single function capturing the singular behavior
of the squared matrix element in both the soft and collinear limits.
Of course, since it is at the level of the amplitude squared
rather than the amplitude, and since it does not take advantage
of the color decomposition, it is not quite a true factorization but is
still tangled up with the color algebra.  The purpose of this note
is to derive a single function
unifying the soft and collinear factors in eqn.~(\use\Factorization)
at the level of color-ordered amplitudes.  This will provide
a true factorization.  I will discuss its connection with 
the Catani--Seymour dipole factorization formula later.
The derivations in ref.~[\use\ManganoReview] treat the soft and 
collinear limits independently.

\section{Derivation of a Unifying Factorization Formula}
\vskip 5pt

Before embarking upon the derivation, I note that
the Koba--Nielsen form is valid whatever the dimensions of the
external momenta and polarization vectors, so long as
the former are on-shell, and the latter transverse.  It thus
allows for a treatment corresponding to any of the possible
variants of dimensional regularization: the conventional
scheme (CDR)~[\use\CDR], 
the original 't~Hooft--Veltman scheme (HV)~[\use\HV], dimensional
reduction (DR)~[\use\DimRed], 
or the four-dimensional helicity scheme (FDH)~[\use\FDHS].

The forms of the collinear and soft factorization functions suggest
that we can associate the singular limits with a trio ($a,1,b$) 
of color-ordered
partons: either a soft parton $1$ with two hard neighbors $(a,b)$, or a pair
of collinear partons neighboring another hard parton, $a\parallel 1$ next
to $b$ or $1\parallel b$ next to $a$.  This in
turn suggests generalizing the derivation in ref.~[\use\ManganoReview]
to extract all singular behavior in either of the two invariants
$s_{a1}$ {\it or\/} $s_{1b}$.  This will summarize the singular
behavior whenever the trio of momenta become degenerate without
either $a$ or $b$ becoming soft, that is as 
$\Delta(a,1,b)/s_{ab}^3\rightarrow0$.  ($\Delta(a,1,b)$ is the
Gram determinant of the ($a,1,b$) system.)  In contrast, the
derivation of the soft limit in ref.~[\use\ManganoReview] takes
both $s_{a1}$ and $s_{1b}$ to be small, and the derivation of
the collinear limit takes only one invariant (say $s_{a1}$) to be small.
The derivation below otherwise follows the same general lines: study
the singular behavior of an amplitude, and identify the parts of the
singular expansion that give rise to an amplitude with fewer external
legs.

Let us then evaluate equation~(\use\KobaNielsen), focussing first on
the singular behavior in either of the two invariants $s_{a1}$ or $s_{1b}$.
Within this singular limit, we may also simplify expressions by expanding
around the field-theory limit $\tau\rightarrow 0$.
If we examine the equation, we see that singularities
in these two invariants can arise only from the integration region
$x_1\simeq x_a=0$ or $x_1\simeq x_b=1$, and then only in those terms
which have a lone inverse power of $x_1$ or $1-x_1$ coming from the
expansion of the exponential.  (The integrals of $x^{-2+\delta}$ 
or $(1-x)^{-2+\delta}$ are finite by analytic continuation as $\delta\rightarrow0$.)
Separating out all terms containing the integration variable $x_1$,
we obtain
$$\eqalign{
&\tau^{n-4}\, 2^{-n/2}\mskip -50mu
\mathop{\int}_{x_a=0<x_1<x_b=1\cdots<x_{n-2}} \prod_{i=1}^{n-3} dx_i
   \,(x_b-x_a)(x_{n-2}-x_a)(x_{n-2}-x_b)\;\cr
&\hskip 20mm\times
 x_1^{\tau^2 k_a\cdot k_1} (1-x_1)^{\tau^2 k_b\cdot k_1}
  \,\prod_{j \in \{2,\ldots,n-2\}} (x_j-x_1)^{\tau^2 k_1\cdot k_j}\;\cr
&\hskip 20mm\times
\left\{\exp\LB 
{\pol_a\tcdot\pol_1\over x_1^2}
+{\pol_b\tcdot\pol_1\over (1-x_1)^2}
-{\tau k_a\tcdot \pol_1\over x_1}
+{\tau k_b\tcdot \pol_1\over (1-x_1)}
+{\tau k_1\tcdot \pol_a\over x_1}
-{\tau k_1\tcdot \pol_b\over (1-x_1)}
\vphantom{\sum_{i \in \{2,\ldots,n-2\}} }\RP\RP\cr
&\hskip 40mm \LP\LP
+\mskip -10mu\sum_{i \in \{2,\ldots,n-2\}} 
\L{\pol_i\tcdot\pol_1\over (x_i-x_1)^2}
               + {\tau k_i\tcdot \pol_1\over (x_i-x_1)}
               - {\tau k_1\tcdot \pol_i\over (x_i-x_1)}\R
\RB\RP\cr
&\hskip 20mm\times
\prod_{\scriptstyle
 i\neq j \in H\atop} |x_j-x_i|^{\tau^2 k_i\cdot k_j/2}\;
\LP\LP\exp\LB 
\sum_{i\neq j \in H} {1\over2} {\pol_i\tcdot\pol_j\over (x_j-x_i)^2}
                         + {\tau k_i\tcdot \pol_j\over (x_i-x_j)}\RB
  \right\}\RV_{\rm multilinear}\,,\cr
}\anoneqn$$
where $H=\{a,b,2,\ldots,n-2\}$ denotes the collection of external
legs excluding leg~1.  The terms on the last line form the integrand
of the $(n-1)$-point amplitude, with leg~1 excluded.  These will
carry along with them $(n-5)$ powers of $\tau$, leaving one to
be associated with the $x_1$ integral.  Noting that we
can partial-fraction terms containing inverse powers of both
$x_1$ and $(1-x_1)$, we see that the $x_1$ integral itself can
produce at most two inverse powers of $\tau$. (Each inverse power
of an invariant comes along with two inverse powers of $\tau$, but
the invariants cannot appear together in the denominator of any
given term.)
 Thus terms in the
integrand containing more than a single additional power of
$\tau$ will disappear in the gauge-theory limit.  This disposes
of the $\prod_{j\in \{2,\ldots,n-2\}}$ factors on the second line,
 after expansion in powers
of $x_1$.
Using this observation in expanding the first exponential as well,
in addition
to dropping terms that have no inverse powers of $x_1$ or $(1-x_1)$ or
are not linear in $\pol_1$, we find,
$$\eqalign{
&\tau^{n-4}\, 2^{-n/2}\mskip -50mu
\mathop{\int}_{x_a=0<x_1<x_b=1\cdots<x_{n-2}} \prod_{i=1}^{n-3} dx_i
   \,(x_b-x_a)(x_{n-2}-x_a)(x_{n-2}-x_b)\;\;
 x_1^{\tau^2 k_a\cdot k_1} (1-x_1)^{\tau^2 k_b\cdot k_1}\cr
&\hskip 20mm\times
\left\{\LB
\vphantom{\sum_{i \in \{2,\ldots,n-2\}}}
\LB 1-
\mskip -10mu\smash{\sum_{i \in \{2,\ldots,n-2\}}}
 {\tau k_1\tcdot \pol_i\over (x_i-x_1)}
\RB\,\LB{\pol_a\tcdot\pol_1\over x_1^2}
+{\pol_b\tcdot\pol_1\over (1-x_1)^2}\RB\RP\RP\cr
&\hskip 30mm\LP
-{\tau k_1\tcdot \pol_b\over (1-x_1)}{\pol_a\tcdot\pol_1\over x_1^2}
+{\tau k_1\tcdot \pol_a\over x_1}{\pol_b\tcdot\pol_1\over (1-x_1)^2}
-{\tau k_a\tcdot \pol_1\over x_1}
+{\tau k_b\tcdot \pol_1\over (1-x_1)}
\vphantom{\sum_{i \in \{2,\ldots,n-2\}} }\RP\cr
&\hskip 30mm \LP\LP
+\LB{\tau k_1\tcdot \pol_a\over x_1}
-{\tau k_1\tcdot \pol_b\over (1-x_1)}
\RB \,\sum_{i \in \{2,\ldots,n-2\}} {\pol_i\tcdot\pol_1\over (x_i-x_1)^2}
\RB_{\rm s.p.}
\,\exp\LB-\mskip -10mu\sum_{i \in \{2,\ldots,n-2\}} 
    {\tau k_1\tcdot \pol_i\over (x_i-x_1)}\RB
\RP\cr
&\hskip 20mm\times
\prod_{\scriptstyle
 i\neq j \in H\atop} |x_j-x_i|^{\tau^2 k_i\cdot k_j/2}\;
\LP\LP\exp\LB 
\sum_{i\neq j \in H} {1\over2} {\pol_i\tcdot\pol_j\over (x_j-x_i)^2}
                         + {\tau k_i\tcdot \pol_j\over (x_i-x_j)}\RB
  \right\}\RV_{\rm multilinear}\,.\cr
}\anoneqn$$

The terms with two inverse powers of $x_1$ or $1-x_1$ will not give
rise to poles after the standard analytic continuation in the invariants.
We can thus drop them inside the first set of brackets.  This subtraction
is indicated by the subscript `s.p.'.  (It is equivalent to freezing $x_1$ at
either $0$ or $1$ as appropriate, in the exponential on the penultimate
line.)

Partial-fraction to isolate the singularities in the $x_1$ integral
and expand the non-singular factors as appropriate either around
$x_1\simeq 0$ or $x_1\simeq 1$,
$$\eqalign{
&\tau^{n-4}\, 2^{-n/2}\mskip -50mu
\mathop{\int}_{x_a=0<x_1<x_b=1\cdots<x_{n-2}} \prod_{i=1}^{n-3} dx_i
   \,(x_b-x_a)(x_{n-2}-x_a)(x_{n-2}-x_b)\;\;
 x_1^{\tau^2 k_a\cdot k_1} (1-x_1)^{\tau^2 k_b\cdot k_1}\cr
&\hskip 10mm\times
\left\{\LB
\vphantom{\sum_{i \in \{2,\ldots,n-2\}}}
{1\over x_1^2} 
\LB 1-\tau k_1\tcdot \pol_b
 -\mskip -10mu\smash{\sum_{i \in \{2,\ldots,n-2\}}}
 {\tau k_1\tcdot \pol_i\over x_i}
\RB\,{\pol_a\tcdot\pol_1}\RP\RP\cr
&\hskip 20mm \vphantom{\sum_{i \in \{2,\ldots,n-2\}}}
+{1\over (1-x_1)^2} 
\LB 1+\tau k_1\tcdot \pol_a-\mskip -10mu\smash{\sum_{i \in \{2,\ldots,n-2\}}}
 {\tau k_1\tcdot \pol_i\over (x_i-1)}
\RB\,{\pol_b\tcdot\pol_1}\cr
&\hskip 20mm\vphantom{\sum_{i \in \{2,\ldots,n-2\}}}
+{1\over x_1}\L\LB 
-\tau k_1\tcdot \pol_b
-\mskip -10mu\smash{\sum_{i \in \{2,\ldots,n-2\}}}
 {\tau k_1\tcdot \pol_i\over x_i^2}
\RB\,\pol_a\tcdot\pol_1
+\tau k_1\tcdot \pol_a\,\pol_b\tcdot\pol_1
-\tau k_a\tcdot \pol_1\RP\cr
&\hskip 40mm\vphantom{\sum_{i \in \{2,\ldots,n-2\}}}\LP
+\tau k_1\tcdot \pol_a\mskip -5mu
   \smash{\sum_{i \in \{2,\ldots,n-2\}}} {\pol_i\tcdot\pol_1\over x_i^2}
\R\cr
&\hskip 20mm\vphantom{\sum_{i \in \{2,\ldots,n-2\}}}
+{1\over (1-x_1)}\L\LB 
+\tau k_1\tcdot \pol_a+\mskip -10mu\smash{\sum_{i \in \{2,\ldots,n-2\}}}
 {\tau k_1\tcdot \pol_i\over (x_i-1)^2}
\RB\,\pol_b\tcdot\pol_1
-\tau k_1\tcdot \pol_b\,\pol_a\tcdot\pol_1
+\tau k_b\tcdot \pol_1\RP\cr
&\hskip 40mm\LP\vphantom{\sum_{i \in \{2,\ldots,n-2\}}}\LP
-\tau k_1\tcdot \pol_b\mskip -5mu
   \smash{\sum_{i \in \{2,\ldots,n-2\}}} {\pol_i\tcdot\pol_1\over (x_i-1)^2}
\R
\RB
\,\exp\LB-\mskip -10mu\sum_{i \in \{2,\ldots,n-2\}} 
    {\tau k_1\tcdot \pol_i\over (x_i-x_1)}\RB
\cr
&\hskip 10mm\times
\prod_{\scriptstyle
 i\neq j \in H\atop} |x_j-x_i|^{\tau^2 k_i\cdot k_j/2}\;
\LP\LP\exp\LB 
\sum_{i\neq j \in H} {1\over2} {\pol_i\tcdot\pol_j\over (x_j-x_i)^2}
                         + {\tau k_i\tcdot \pol_j\over (x_i-x_j)}\RB
  \right\}\RV_{\rm multilinear}\,.\cr
}\anoneqn$$

It is also convenient at this stage to extract
all the terms linear in $\pol_a$ or $\pol_b$ 
from the exponential on the last line, whence we obtain
$$\eqalign{
&\tau^{n-4}\, 2^{-n/2}\mskip -50mu
\mathop{\int}_{x_a=0<x_1<x_b=1\cdots<x_{n-2}} \prod_{i=1}^{n-3} dx_i
   \,(x_b-x_a)(x_{n-2}-x_a)(x_{n-2}-x_b)\;\;\tau\;
 x_1^{\tau^2 k_a\cdot k_1} (1-x_1)^{\tau^2 k_b\cdot k_1}\cr
&\hskip 10mm\times
\left\{\LB
\vphantom{\sum_{i \in \{2,\ldots,n-2\}}}
{1\over x_1}\L\LB 
- k_1\tcdot \pol_b
-\mskip -15mu\smash{\sum_{i\neq j \in \{2,\ldots,n-2\}}}
 { k_1\tcdot \pol_i\,\pol_b\tcdot\pol_j\over x_i^2 (x_j-1)^2}
\RB\,\pol_a\tcdot\pol_1
+ k_1\tcdot \pol_a\,\pol_b\tcdot\pol_1
- k_a\tcdot \pol_1\,\pol_a\tcdot\pol_b
\RP\RP\RP\cr
&\hskip 30mm\vphantom{\sum_{i \in \{2,\ldots,n-2\}}}\LP
- k_a\tcdot \pol_1\mskip -5mu\smash{\sum_{i\neq j \in \{2,\ldots,n-2\}}}
       {\pol_a\tcdot\pol_i\,\pol_b\tcdot\pol_j\over x_i^2 (x_j-1)^2}
+ k_1\tcdot \pol_a\mskip -5mu
   \smash{\sum_{i\neq j \in \{2,\ldots,n-2\}}} 
     {\pol_i\tcdot\pol_1\,\pol_b\tcdot\pol_j\over x_i^2 (x_j-1)^2}
\R\cr
&\hskip 20mm\vphantom{\sum_{i \in \{2,\ldots,n-2\}}}
+{1\over (1-x_1)}\L\LB 
k_1\tcdot \pol_a+\mskip -15mu\smash{\sum_{i \in \{2,\ldots,n-2\}}}
 { k_1\tcdot \pol_i\,\pol_a\tcdot\pol_j\over (x_i-1)^2 x_j^2}
\RB\,\pol_b\tcdot\pol_1
- k_1\tcdot \pol_b\,\pol_a\tcdot\pol_1
+ k_b\tcdot \pol_1\,\pol_a\tcdot\pol_b
\RP\cr
&\hskip 40mm\LP\vphantom{\sum_{i \in \{2,\ldots,n-2\}}}\LP
+ k_b\tcdot \pol_1\mskip -5mu\smash{\sum_{i\neq j \in \{2,\ldots,n-2\}}}
       {\pol_a\tcdot\pol_i\,\pol_b\tcdot\pol_j\over \smash{x_i^2 (x_j-1)^2} 
       \vphantom{x_i}}
- k_1\tcdot \pol_b\mskip -5mu
   \smash{\sum_{i \in \{2,\ldots,n-2\}}} 
     \smash{{\pol_i\tcdot\pol_1\,\pol_a\tcdot\pol_j\over (x_i-1)^2 x_j^2}}
\R
\RB\cr
&\hskip 20mm\times\exp\LB-\mskip -10mu\sum_{i \in \{2,\ldots,n-2\}} 
    \L{\tau k_1\tcdot \pol_i\over (x_i-x_1)}
+ {\tau k_a\tcdot \pol_j\over x_i}
+ {\tau k_b\tcdot \pol_j\over (x_i-1)}\R
\RB
\cr
&\hskip 10mm\times
\prod_{\scriptstyle
 i\neq j \in H\atop} |x_j-x_i|^{\tau^2 k_i\cdot k_j/2}\;
\LP\LP\exp\LB 
\sum_{i\neq j \in \{2,\ldots,n-2\}} 
{1\over2} {\pol_i\tcdot\pol_j\over (x_j-x_i)^2}
                         + {\tau k_i\tcdot \pol_j\over (x_i-x_j)}\RB
  \right\}\RV_{\rm multilinear}\,.\cr
}\anoneqn$$

\def\ah{{\hat a}}
\def\bh{{\hat b}}
\def\polstar{\pol{\vphantom{\LB K\RB}}}

Next, introduce polarization vectors $\pol_\ah,\pol_\bh$
for two new momenta $k_\ah, k_\bh$, and
replace two dot products in each term via the identity
$$
g^{\mu\nu} = -\mskip -20mu\sum_{{\rm polarizations\ } h} \mskip -10mu
   \pol^\mu_{(h)} \pol^{\nu*\vphantom{\mu}}_{(h)}
\anoneqn$$
where the sum runs over both physical and unphysical polarizations.
\def\H{{\widehat H}}

The values of $k_\ah, k_\bh$ will be fixed later; for the moment, they
are arbitrary.
Collecting terms, we obtain
$$\eqalign{
&\tau^{n-4}\, 2^{-n/2}\mskip -50mu
\mathop{\int}_{x_a=0<x_1<x_b=1\cdots<x_{n-2}} \prod_{i=1}^{n-3} dx_i
   \,(x_b-x_a)(x_{n-2}-x_a)(x_{n-2}-x_b)\;\;\tau\;
 x_1^{\tau^2 k_a\cdot k_1} (1-x_1)^{\tau^2 k_b\cdot k_1}\cr
&\hskip 7mm\times \mskip -40mu\sum_{{\rm polarizations}(\ah,\bh)}
\left\{\LB
\vphantom{\sum_{i \in \{2,\ldots,n-2\}}}
{1\over x_1}\pol_\bh\tcdot\pol_b \,
\L \vphantom{\sum}
  -\pol_a\tcdot\pol_1\,k_1\tcdot\pol_\ah\, 
+ k_1\tcdot \pol_a\, \pol_\ah\tcdot\pol_1\,
- k_a\tcdot \pol_1\,\pol_a\tcdot\pol_\ah 
\R\RP\RP\cr
&\hskip 22mm\vphantom{\sum_{i \in \{2,\ldots,n-2\}}}
\times\LB
 \polstar_\ah^*\tcdot\polstar_\bh^*
+\mskip -15mu\smash{\sum_{i\neq j \in \{2,\ldots,n-2\}}}
 { \polstar_\ah^*\tcdot\pol_i\,
       \polstar_\bh^*\tcdot\pol_j\over x_i^2 (x_j-1)^2}\RB
\,\exp\LB-\mskip -10mu\sum_{i \in \{2,\ldots,n-2\}} 
    \L{\tau (k_a+k_1)\tcdot \pol_i\over x_i}
    + {\tau k_b\tcdot \pol_j\over (x_i-1)}\R\RB
\cr
&\hskip 15mm\vphantom{\sum_{i \in \{2,\ldots,n-2\}}}
+{1\over (1-x_1)}\pol_\ah\tcdot \pol_a\,
\L \vphantom{\sum}\pol_b\tcdot\pol_1\,k_1\tcdot\pol_\bh
- k_1\tcdot \pol_b\,\pol_\bh\tcdot \pol_1
+ k_b\tcdot \pol_1\,\pol_b\tcdot\pol_\bh\R\cr
&\hskip 22mm\LP\vphantom{\sum_{i \in \{2,\ldots,n-2\}}}
\times\LB
 \polstar_\ah^*\tcdot \polstar_\bh^*
+ \mskip -15mu\smash{\sum_{i\neq j \in \{2,\ldots,n-2\}}}
       {\polstar_\ah^*\tcdot \pol_i\,\polstar_\bh^*\tcdot \pol_j
         \over \smash{x_i^2 (x_j-1)^2} 
       \vphantom{x_i}}\RB
\,\exp\LB-\mskip -10mu\sum_{i \in \{2,\ldots,n-2\}} 
    \L{\tau k_a\tcdot \pol_j\over x_i}
+ {\tau (k_b+k_1)\tcdot \pol_j\over (x_i-1)}\R
\RB
\RB\cr
&\hskip 10mm\times
\prod_{\scriptstyle
 i\neq j \in H\atop} |x_j-x_i|^{\tau^2 k_i\cdot k_j/2}\;
\LP\LP\exp\LB 
\sum_{i\neq j \in \{2,\ldots,n-2\}} 
   {1\over2} {\pol_i\tcdot\pol_j\over (x_j-x_i)^2}
                         + {\tau k_i\tcdot \pol_j\over (x_i-x_j)}\RB
  \right\}\RV_{\rm multilinear}\,,\cr
}\anoneqn$$
where `multilinear' is now understood to refer to $\pol$ and
$\pol^*$ independently, and where I have set $x_1$ in the exponential
to $0$ or $1$ as appropriate to the pole structure of each term.

Take $x_\ah = 0$, $x_\bh=1$, and define
$\H = \{-\ah,-\bh,2,\ldots,n-2\}$; `$-\ah$' signifies the replacement
of $(k_\ah,\polstar_\ah^*)$ by $(k_{-\ah},\pol_{-\ah})\equiv(-k_{\ah},\pol_{\ah}^*)$.
We can absorb the factors in brackets
into the exponential on the last line,
$$\eqalign{
&\tau^{n-4}\, 2^{-n/2}\mskip -50mu
\mathop{\int}_{x_a=0<x_1<x_b=1\cdots<x_{n-2}} \prod_{i=1}^{n-3} dx_i
   \,(x_b-x_a)(x_{n-2}-x_a)(x_{n-2}-x_b)\;\;\tau\;
 x_1^{\tau^2 k_a\cdot k_1} (1-x_1)^{\tau^2 k_b\cdot k_1}\cr
&\hskip 10mm\times \mskip -40mu\sum_{{\rm polarizations}(\ah,\bh)}
\left\{\LB
\vphantom{\sum_{i \in \{2,\ldots,n-2\}}}
{1\over x_1}\pol_\bh\tcdot\pol_b \,
\L \vphantom{\sum}
  -\pol_a\tcdot\pol_1\,k_1\tcdot\pol_\ah\, 
+ k_1\tcdot \pol_a\, \pol_\ah\tcdot\pol_1\,
- k_a\tcdot \pol_1\,\pol_a\tcdot\pol_\ah
\R\RP\RP\cr
&\hskip 30mm\vphantom{\sum_{i \in \{2,\ldots,n-2\}}}
\times\exp\LB-\mskip -10mu\sum_{i \in \{2,\ldots,n-2\}} 
    \L{\tau (k_a+k_1)\tcdot \pol_i\over x_i}
    + {\tau k_b\tcdot \pol_j\over (x_i-1)}
    +{\tau k_\ah\tcdot \pol_i\over x_i}
    +{\tau k_\bh\tcdot \pol_j\over (x_i-1)}\R\RB
\cr
&\hskip 20mm\vphantom{\sum_{i \in \{2,\ldots,n-2\}}}
+{1\over (1-x_1)}\pol_\ah\tcdot \pol_a\,
\L\vphantom{\sum} \pol_b\tcdot\pol_1\,k_1\tcdot\pol_\bh
- k_1\tcdot \pol_b\,\pol_\bh\tcdot \pol_1
+ k_b\tcdot \pol_1\,\pol_b\tcdot\pol_\bh\R\cr
&\hskip 30mm\LP\vphantom{\sum_{i \in \{2,\ldots,n-2\}}}
\times\exp\LB-\mskip -10mu\sum_{i \in \{2,\ldots,n-2\}} 
    \L{\tau k_a\tcdot \pol_j\over x_i}
      + {\tau (k_b+k_1)\tcdot \pol_j\over (x_i-1)}
      +{\tau k_\ah\tcdot \pol_j\over x_i}
      +{\tau k_\bh\tcdot \pol_j\over (x_i-1)}\R\RB
\RB\cr
&\hskip 10mm\times
\prod_{\scriptstyle
 i\neq j \in H\atop} |x_j-x_i|^{\tau^2 k_i\cdot k_j/2}\;
\LP\LP\exp\LB 
\sum_{i\neq j \in \H} 
   {1\over2} {\pol_i\tcdot\pol_j\over (x_j-x_i)^2}
                         + {\tau k_i\tcdot \pol_j\over (x_i-x_j)}\RB
  \right\}\RV_{\rm multilinear}\,,\cr
}\anoneqn$$
up to terms which vanish in the gauge-theory limit.

If we now choose $k_\ah = -k_a-k_1$ and $k_\bh = -k_b$ in the 
first term, and $k_\ah = -k_a$ and $k_\bh = -k_b-k_1$ in the 
second, we can perform the $x_2\ldots x_{n-3}$ integrations.
They yield an $(n-1)$-point amplitude, so that the singular
limit of the original amplitude becomes
$$\eqalign{
&{\tau^2\over \sqrt{2}}\int_0^1 dx_1\;
 x_1^{\tau^2 k_a\cdot k_1} (1-x_1)^{\tau^2 k_b\cdot k_1}\cr
&\hskip 10mm\times \mskip -40mu\sum_{{\rm polarizations}(\ah,\bh)}
\LB
\vphantom{\sum_{i \in \{2,\ldots,n-2\}}}
{1\over x_1}\pol_\bh\tcdot\pol_b \,
\L \vphantom{\sum}
  -\pol_a\tcdot\pol_1\,k_1\tcdot\pol_\ah\, 
+ k_1\tcdot \pol_a\, \pol_\ah\tcdot\pol_1\,
- k_a\tcdot \pol_1\,\pol_a\tcdot\pol_\ah
\R\RP\cr
&\hskip 40mm\vphantom{\sum_{i \in \{2,\ldots,n-2\}}}
\times A_{n-1}^{\rm KN}(\ldots,-\ah=a+k_1,-\bh=b,\ldots)
\cr
&\hskip 30mm\vphantom{\sum_{i \in \{2,\ldots,n-2\}}}
+{1\over (1-x_1)}\pol_\ah\tcdot \pol_a\,
\L \vphantom{\sum}\pol_b\tcdot\pol_1\,k_1\tcdot\pol_\bh
- k_1\tcdot \pol_b\,\pol_\bh\tcdot \pol_1
+ k_b\tcdot \pol_1\,\pol_b\tcdot\pol_\bh\R\cr
&\hskip 40mm\LP\vphantom{\sum_{i \in \{2,\ldots,n-2\}}}
\times A_{n-1}^{\rm KN}(\ldots,-\ah=a,-\bh=b+k_1,\ldots)
\RB\cr
}\eqn\FinalSingular$$

Performing the $x_1$ integration and taking the $\tau\rightarrow 0$
limit then gives us 
$$\eqalign{
&\mskip -40mu\sum_{{\rm polarizations}(\ah,\bh)}
\LB
{1\over s_{a1}}\pol_\bh\tcdot\pol_b \,V_3(a,1,\ah)
\, A_{n-1}^\tree(\ldots,-\ah=a+k_1,-\bh=b,\ldots)
\RP\cr
&\hskip 30mm\LP
-{1\over s_{b1}}\pol_\ah\tcdot \pol_a\,V_3(b,1,\bh)
\, A_{n-1}^\tree(\ldots,-\ah=a,-\bh=b+k_1,\ldots)
\RB\cr
}\eqn\FinalSingularB$$
where $V_3(a,b,c)$ is a certain form of
the color-ordered three-point vertex (up to
a factor of $i$),
$$
V_3(1,2,3) = 
\sqrt2 \L  -\pol_1\tcdot\pol_2\,k_2\tcdot\pol_3\, 
+ k_2\tcdot \pol_1\, \pol_3\tcdot\pol_2\,
- k_1\tcdot \pol_2\,\pol_1\tcdot\pol_3\R\,.
\anoneqn$$
  In the singular limit, one might
write down other forms of the three-point vertex.  However, later
formul\ae\ will not conserve momentum amongst the three momentum
in these vertices (they will of course conserve momentum overall),
and as we will be taking the ratio of two quantities which both
vanish in the singular limit, we
must be careful which form we use.  (With this in mind, one could
also use the recurrence relations formalism~[\use\Recurrence]
to derive eqn.~(\use\FinalSingularB).)

If we can find a pair of {\it reconstruction\/}
functions $k_{\ah,\bh} = f_{\ah,\bh}(k_a,k_1,k_b)$
such that $(k_\ah,k_\bh)\rightarrow -(k_a+k_1,k_b)$ sufficiently quickly
when $k_1\parallel k_a$,
and likewise $(k_\ah,k_\bh)\rightarrow -(k_a,k_b+k_1)$ 
sufficiently quickly when $k_1\parallel k_b$,
then we can pull out a common factor of the surviving hard amplitude.
Furthermore, gauge
invariance allows us to replace the sum over polarizations with a sum over helicities,
so we obtain for the singular limit,
$$\eqalign{
&\mskip -40mu\sum_{{\rm helicities}(\ah,\bh)}
\LB
{1\over s_{a1}}\pol_\bh\tcdot\pol_b \,V_3(a,1,\ah)
-{1\over s_{b1}}\pol_\ah\tcdot \pol_a\,V_3(b,1,\bh)
\RB
\, A_{n-1}^\tree(\ldots,-\ah(a,1,b),-\bh(b,1,a),\ldots)\,.
\cr}\anoneqn$$
(In the four-dimensional helicity scheme, only the $\pm$ helicities are needed;
in the conventional and 't~Hooft--Veltman schemes, `$\e$' helicities are needed as
well.)  The factor in brackets is the {\it antenna\/} factorization function,
$$
\Sing(\ah,\bh\longleftarrow a,1,b) = 
{1\over s_{a1}}\pol_\bh\tcdot\pol_b \,V_3(a,1,\ah)
-{1\over s_{b1}}\pol_\ah\tcdot \pol_a\,V_3(b,1,\bh)\,.
\eqn\AntFactor$$

This function factorizes the singular region of three-particle
production onto the antenna formed by two hard particles.
It depends on the helicities of the original particles, as well
as the helicities of the resulting hard particles.

\section{Explicit Forms}
\vskip 5pt

In the $k_a\parallel k_1$ 
collinear limit, the second term in eqn.~(\use\AntFactor)
is non-singular, and we recover the usual collinear splitting amplitude,
given by the first term~[\use\ManganoReview] (note that the normalization
conventions there are slightly different from those in the present paper).
Similarly, in the $k_b\parallel k_1$ limit, only the second term contributes.
In the soft limit ($k_1\rightarrow 0$),
$k_\ah=-k_a$, $k_\bh=-k_b$, and we can neglect all
terms proportional to $k_1$ in $V_3$; we then obtain
$$
\pol_\ah\tcdot \pol_a\,\pol_b\tcdot\pol_\bh\,
\LB -{1\over s_{a1}} k_a\tcdot \pol_1
+ {1\over s_{b1}} k_b\tcdot \pol_1\RB\,,
\anoneqn$$
which is just the usual eikonal factor.  Thus we recover known factorization
formul\ae\ in the appropriate subregions.

Sensible reconstruction functions satisfy momentum conservation
everywhere, and also keep $k_{\ah,\bh}$ massless everywhere (not just in
the singular limit).  This corresponds to taking the sum of the original
momenta, $K=k_a+k_1+k_b$, and rewriting it as the sum of two massless momenta.
Working in the frame where ${\bf K} = 0$, we see that there is a degeneracy
corresponding to the relative angle of ${\bf b}$ and $-{\bf \bh}$.  In the
singular limit, this degeneracy is fixed by the requirement that the angle
vanish; but away from the limit, it survives.  This means that the choice
of reconstruction function is not unique.

One possibility is to use the Catani--Seymour choice,
$$
k_\ah = -k_a - k_1 + {s_{a1}\over s_{1b}+s_{ab}} k_b,\hskip 10mm
k_\bh = -{K^2\over s_{1b}+s_{ab}}\, k_b\,,
\anoneqn$$
which however is asymmetric in our case (and is not valid for $k_b\parallel k_1$).
  A more symmetric choice is
to take
$$\eqalign{
s_{\ah a} &= -{s_{a1} s_{1b}\over s_{1b}+s_{ab}} r_1 \LB 1
              - {2 (1-r_1)\over\rho+1} \RB
\,,\cr
s_{\bh b} &= -{s_{a1} s_{1b}\over s_{1a}+s_{ab}} (1-r_1) \LB 1
              - {2 r_1\over\rho+1} \RB
\,,\cr
}\anoneqn$$
where
$$
r_1 = {s_{1b}\over s_{1a}+s_{1b}}\,,\hskip 10mm
\rho = \sqrt{1 + {4 r_1 (1-r_1) s_{1a} s_{1b}\over K^2 s_{ab}}}\,.
\anoneqn$$
(Note that $\rho\rightarrow 1$ in all singular limits.)
This is equivalent to
$$\eqalign{
k_\ah &= -{1\over 2} \LB 1+\rho + {s_{1b} (1+\rho-2 r_1)\over s_{1a}+s_{ab}}\RB\, k_a
          - r_1 k_1 
    -{1\over 2}\LB 1-\rho + {s_{1a} (1-\rho-2 r_1)\over s_{1b}+s_{ab}}\RB\, k_b\,,\cr
k_\bh &= -{1\over 2} \LB 1-\rho - {s_{1b} (1+\rho-2 r_1)\over s_{1a}+s_{ab}}\RB\, k_a
          - (1-r_1) k_1 
    -{1\over 2}\LB 1+\rho - {s_{1a} (1-\rho-2 r_1)\over s_{1b}+s_{ab}}\RB\, k_b\,,\cr
}\anoneqn$$
These approach the appropriate Catani--Seymour choice in both
collinear limits, $a\parallel 1$ and $b\parallel 1$.

So long as we stay in four dimensions, we can use the spinor
helicity method to give a concrete expression for the antenna
function.
Given values for the momenta $k_i$, these quantities can
be evaluated numerically.  For example,
$$\eqalign{
\Sing(\ah^-,\bh^-\longleftarrow a^+,1^+,b^+) &= 
-{K^2\over \spa{a}.1\spa1.b} 
 {\spb{a}.b \spa{b}.{\ah}\spa{a}.{\smash{\bh}} \over
  \spa{a}.b^2 \spb{b}.{\ah} \spb{a}.{\smash{\bh}}}\,,
}\anoneqn$$
where I have used identities such as
$$
{\spa{q}.{\ah}\over\spa{q}.a} = {\spa{b}.{\ah}\over\spa{b}.a}
 + {\spa{q}.b\spa{\ah}.{a}\over \spa{q}.a\spa{b}.a}\,,
\anoneqn$$
and have dropped terms such as the second which are non-singular
in the limit.

Squaring the antenna function, summing over final helicities ($a,1,b$),
and averaging over initial helicities ($\ah,\bh$), one obtains
$$
2 {\LB K^2 (K^2-s_{ab})+s_{ab}^2\RB^2\over s_{a1} s_{1b} s_{ab} (K^2)^2} + \cdots\,,
\eqn\AntennaSquared$$
where the omitted terms are less singular, and in particular lead to
finite integrals (as $\e\rightarrow0$) 
when integrated over `singular' (soft or collinear) regions
of phase space.  (The inclusion or exclusion of such terms is arbitrary, as
they are not universal.  In practical applications one may choose to omit
them.)
The reader may verify that this expression reproduces
known expressions in the soft and collinear limits.

Away from four dimensions, we must take the unobserved momenta ($a,1,b$)
to have $4-2\e$ rather than four components; depending on the variant
of dimensional regularization, we may also take polarization vectors
or observed momenta ($\ah,\bh$) to have $4-2\e$ rather than four components.
In the CDR scheme, one obtains
the same expression~(\use\AntennaSquared) for the antenna function;
in other schemes, there will in general be additional contributions.

If instead we choose the Catani--Seymour reconstruction function, we cannot
expect to reproduce the correct behavior in the $k_b\parallel k_1$ limit, but
we will reproduce the Catani--Seymour dipole factorization in
both the $k_a\parallel k_1$ and
$k_1\rightarrow 0$ limits.  (Although the antenna function is not
equal to the Catani--Seymour dipole function, the difference between them
consists of terms which are not singular in the limit.  Such terms,
as noted above, are not universal, and may be omitted at will.)
The corresponding helicity amplitudes could be thought of
as the `square root' of the dipole function.
The antenna function may be thought of as a
generalization of the dipole factorization formula.

Although I have discussed only the antenna factorization function for
purely gluonic amplitudes, it is clear that one can write down
analogous formulae for mixed quark-gluon amplitudes.  The approach presented
here has a straightforward generalization to multiple-singular emission, which
I shall endeavor to discuss elsewhere.  I thank Z. Bern, S. Catani, L. Dixon,
and N. Glover for helpful comments.

\listrefs
\bye